\documentclass[nohyper,12pt,letterpaper]{JHEP}
\usepackage{epsfig}

\newfont{\frak}{eufm10 scaled 1200}

\newfont{\Bbb}{msbm10 scaled 1200}     
\newcommand{\mathbb}[1]{\mbox{\Bbb #1}}
\DeclareSymbolFont{AMSa}{U}{msa}{m}{n}
\DeclareSymbolFont{AMSb}{U}{msb}{m}{n}
\let\Box\relax
\DeclareMathSymbol{\Box}{\mathord}{AMSa}{"03}

\def \eqn#1#2{\begin{equation}#2\label{#1}\end{equation}}

\title{An Holographic Cosmology}

\author{T. Banks${}^*$\\\thanks{On Leave from Dept. of Physics, Rutgers U.}
   Department of Physics and Institute for Particle Physics\\
   University of California, Santa Cruz, CA 95064\\
E-mail: \email{banks@scipp.ucsc.edu}}

\author{W.Fischler${}^{**}$\\
Department of Physics and Astronomy\\ University of Texas, Austin,
TX
\\ E-mail: \email{fischler@physics.utexas.edu}}

\abstract{We present a new cosmological model, based on the holographic
principle, which shares many of the virtues of inflation.  The very
earliest semiclassical era of the universe is dominated by a dense
gas of black holes, with equation of state $p = \rho$.
Fluctuations lead to an instability to a phase with a dilute gas
of black holes, which later decays via Hawking radiation to a
radiation dominated universe.  The quantum fluctuations of the
initial state give rise to a scale invariant spectrum of density
perturbations, for a range of scales.  We point out a problem
, that appears to prevent the range of scales predicted by the model
from coinciding with the range where such a spectrum has been observed.
We speculate that this may be related to our field theoretic treatment
of fluctuations in the highly holographic $p=\rho$ background. The
monopole problem is solved in a manner completely different from inflationary
models, and a relic density of highly charged extremal black monopoles is
predicted. We discuss the nature of the entropy and flatness problems
in our model. }

\keywords{Cosmology, Holographic Principle}

\received{???????? ?st, 2000} \accepted{???????? ?th, 1998}
\preprint{\hepth{0111142}\\RUNHETC-2001-22\\SCIPP-00/21\\UTTG-13-01}

\begin{document}




\section{\bf Introduction}


In a recent paper\cite{bf} we presented an introduction to a
quantum theory of cosmology.  The focus of this as yet incomplete
formalism was the combination of the holographic principle with
the conventional notion of particle horizon.  The fact that these
two concepts are compatible with each other was an important ,
though implicit, message of \cite{fs}.  We used insights from
this formalism to begin the construction of a semiclassical
cosmology, which would be compatible with the basic principles of
M-theory.  We argued that the earliest semiclassical era of the
universe consisted of a homogeneous gas with equation of state $p
= \rho$, which saturates the F(ischler)-S(usskind)-B(ousso)
entropy bound.  We pointed out that this provided a
noninflationary solution of the horizon problem.  That is, a
homogeneous gas can carry the entire entropy allowed by the
holographic bound, so there are no significant inhomogeneous
perturbations of the system\footnote{In a forthcoming paper \cite{bc} we
show that a $p=\rho$ fluid cannot support growing inhomogeneities, even
near a Big Crunch singularity.}.

In this paper we show that a dense gas of black holes, one in
which the typical black hole size is of order the particle
horizon, and the typical distance between black holes is of order
their Schwarzchild radius, provides a picturesque physical
realization of the $p=\rho$ gas.  This description allows us to
obtain a crude picture of the evolution of the universe.  We argue
that, apart from quantum and statistical fluctuations, the
$p=\rho$ state is stable.  Black holes merge to form a single
horizon sized black hole as the universe expands.  However,
quantum mechanically, there is a nonzero probability that some
horizon volumes have smaller than average black holes in them.  If
the black holes in a sufficient number of contiguous horizon
volumes are small enough, they cease to merge, and evolve as a
$p=0$ gas.  The primordial black holes decay rapidly and produce
small radiation dominated regions. These regions grow in physical
volume, more rapidly than the $p=\rho$ background and after some
time the universe consists of large $p={1\over 3} \rho$ regions
with black holes (the remnants of the $p=\rho$ phase)
caught in the interstices between them. The
coarse grained universe is thus described by a nonrelativistic gas
of large black holes. There is also likely to be a large $p=\rho$
region outside the combined particle horizons of the 
$p=\rho /3$ region.
We will refer to the $p=\rho$ regime as the dense black hole
fluid, and the $p=\rho /3$ regime as the dilute black hole gas.

The dilute black hole gas regions evolve like a conventional
matter dominated cosmology. However, the black holes which
comprise the non relativistic matter are unstable to decay via
Hawking radiation. When these black holes decay
the universe becomes radiation dominated.  The largest black
holes are those caught in the interstices between different $p=\rho /3$
regions, rather than incorporated within any one original
region.  These dominate the energy density until the time they
decay.

The fluctuations in the distribution of matter in the $p=\rho /3$
regions are thus the fluctuations in the distribution of these
largest black holes.  We argue that at very large length scales,
these fluctuations are determined by the quantum fluctuations in
the original distribution of $p=\rho /3$ regions in the $p=\rho$ gas. We
make the hypothesis that the form of these quantum fluctuations at
large scale is calculable in terms of quantum field theory and
comes predominantly from the long range two point correlations of
the gravitational field . The large scale fluctuations are thus
Gaussian. We further show that in the $p=\rho$ background, they
are scale invariant: the magnitude of fluctuations at the horizon
scale is independent of how large the horizon scale is. Standard
arguments show that once these quantum fluctuations have been
converted into classical matter density fluctuations, this scale
invariance persists and will show up in the cosmic microwave
spectrum.

Two crucial assumptions in the above discussion of fluctuations
are that the initial density of dilute black hole gas regions is
small and that the fluctuations of this distribution around its
average are even smaller. We argue that both these facts are
consequence of the low probability $P_0$ of finding dilute gas
regions large enough to destabilize the dense black hole fluid.
The small initial density of dilute gas regions implies that the
dense black hole fluid persists for a considerable time. This has
two important consequences.   It contributes to the solution of
  the monopole problem, by a mechanism adumbrated below, and it
determines the reheat temperature of the universe.  The
fundamental parameter $P_0$ appears to be determined by Planck
scale physics, for reasons we will explain below.

Our quantum cosmology thus provides a mechanism for producing a
homogeneous isotropic universe with a spectrum of small, Gaussian,
scale invariant, fluctuations.  However, we argue that causality
constraints on our quantum field theoretic formalism imply that
the largest scale for which we predict a scale invariant spectrum
is $10^7$ times smaller than the current horizon radius.  This
suggests that (either our model is wrong or) our field theoretic
treatment of the dense black hole gas is an inadequate description
of this highly holographic system.  In the conclusions we will
speculate about how a proper quantum treatment of this regime
might remove the discrepancy that we have found.

  We will argue that our model
provides insight into the other cosmological conundra that
inflation was invented to solve.  Briefly, the monopole problem is
solved because monopoles originate as magnetically charged black
holes.  There is a small entropic suppression of such black holes.
However, the major part of the resolution of the monopole problem
comes from the length of time spent in the $p=\rho$ era.  During
this period, black holes are growing via mergers.  Since the era
lasts for a long time, they are quite massive by the time the
universe becomes matter dominated (for the first time).  They also
develop a very large charge.   The magnetically charged black
holes will remain as remnants and will evolve into extremal
magnetic black holes with huge charge.  We will call these
remnants, and their nonextremal progenitors, {\it black
monopoles}.

Hawking decay of both charged and neutral black holes produces a
large number of photons per monopole.  For a sufficiently long
$p=\rho$ era, the monopole to photon ratio is small enough to
satisfy all known bounds.   In contrast to the inflationary
scenarios, our cosmology might produce an observable relic
density of monopoles.  Furthermore, the relic black monopoles are
exotic and have extremely large charge.

Another problem whose resolution is credited to inflation is
the {\it entropy problem}.  We have often been confused by
statements of this problem in the literature.   Many of them are
mathematically equivalent to the following fact: the linear size
of our present horizon volume, at the era when the energy density
approaches the Planck scale, is $10^{29} $ in Planck units.  In
our view, this large number reflects the holographic bounds and
the total number of quantum states of the universe.

The number of quantum states in the universe is determined by the
cosmological constant.  Any geometrical description of the
universe must then begin with a large spacetime, even at the
Planck density. The precise way in which the available states are
distributed between the cosmological horizon and more localized
systems is determined by dynamics.   The conventional count of
the entropy of the universe includes only the localized
entropy.   In our model, this will be explained by detailed
dynamical calculations.

Finally, we come to the flatness problem, whose resolution is one
of the phenomenological triumphs of inflation.  Our viewpoint on
this is unorthodox.  We are tempted to view the problem as
resulting from an unjustified use of homogeneous isotropic
cosmology to describe the very early universe, an improper
mingling of phenomenology and fundamental theory. Indeed, the
horizon problem tells us that homogeneous isotropic cosmology
should be derived rather than postulated and this is the case in
our model.   We will provide dynamical justification for flatness
within a horizon volume at early times, and we believe that this
is enough to account for observations.  More global requirements
of flatness seem unjustified to us.  Small deviations from the
flat background geometry in each horizon volume have no reason to
be aligned to produce an average curvature.  Rather, they should
be viewed as random fluctuations in spatial curvature, which,
using Einstein's equations, are equivalent to density
fluctuations.  We will calculate these fluctuations in our model,
and they do not give rise to an average curvature.

In the course of our discussion of black monopole evaporation, we
will note that these processes provide an unconventional source
of baryon asymmetry, whose order of magnitude is hard to
calculate.   We will also find indications, though no firm
argument, that the reheat temperature of the universe is very
low, so that many conventional methods of producing baryon
asymmetry will not be applicable.  Affleck-Dine \cite{ad}
coherent baryogenesis would be the leading candidate mechanism,
unless the asymmetry generated in black monopole interactions and
decay is sufficient. We will also make some brief comments about
gravitinos, axions, and dark matter.

To summarize the plan of this paper: Section 2 is the heart of
the paper and fills in the sketch we have presented above.
Section 3 is devoted to a short critique of inflationary models
in view of the nonzero value of the cosmological constant, and a
comparison of such models with our own.  It also contains our
conclusions.

Before proceeding, a word about the relation of our work to
M-theory. This rubric currently applies to a collection of
supersymmetric vacuum states with Poincare or Anti-deSitter
SUSY.   There are plausible arguments for SUSY violating vacua
with AdS symmetry, as well as SUSY violating, weakly coupled,
string cosmologies with no cosmological horizon.  We believe that
all of these can be viewed as limits of a more general structure
whose basic rules will incorporate and make precise the
Holographic Principle\cite{tHS}, and the UV/IR
connection\cite{sw}.  Apart from a brief remark that higher
dimensional physics is important for the determination of the
parameters $P_0$ and $M_0$ defined below, our cosmological model
will employ only these two fundamental principles of M-theory.

\section{\bf Semiclassical cosmology}

The considerations of \cite{bf} motivate us to choose a time
slicing of spacetime such that all backward light cones with their
tips on a given time slice, have the same FSB area.   Once this
area is large enough, we expect to be able to describe much of the
physics in terms of a background spacetime, and localized
fluctuations in it. Let $H_0$ be the Hubble parameter at the first
instant at which we believe semiclassical geometry (but not
quantum field theory) is a good approximation. $H_0^{-1}$ must be
larger than the Planck scale; precisely how large is not clear.
For brevity, we will refer to this first semiclassical time slice
as {\it the initial instant} in the remainder of the text.
Locally, within a backward light cone ending on this time slice,
it makes sense to view the universe as an FRW space with some sort
of localized excitations in it.  This defines a Hamiltonian time
evolution according to the FRW time. The question is, what should
we take for the initial state?  In our previous paper we argued
that this should be a fluid with $p=\rho$, because such a fluid
dominates the energy density at small scale factor, and is the
only homogeneous fluid that can saturate the FSB entropy bounds
for all times. We will begin by choosing a flat FRW space, and
then discuss other options.

It does not make sense to search for low energy states of the FRW
Hamiltonian.  Prior to this time, this particular choice of time
evolution did not make any sense.  It is only well defined in the
semiclassical regime\cite{tbfsrub}.  Rather, we should choose a
generic initial state, consistent with the fact that it is
contained in our backward light cone. Here, the UV/IR connection
of M-theory comes in handy.  It assures us that the density of
states grows rapidly at very high FRW energy, and that the high
energy states all have the geometry of black holes.   We do not
have a good microscopic description of these states from the point
of view of an observer who hovers outside the black hole, but a
loose application of the No-Hair theorem assures us that they all
have the same macroscopic geometry for given values of charge,
mass and angular momentum.

It is important to note that black holes are really only defined
in terms of asymptotic geometry.  We are thinking of black holes
inside a backward light cone as regions bounded by an apparent
horizon of a certain area.  Our discussion of their properties is
only approximate. It is clear what we mean by such a black hole in
the limit that its radius is small compared to the horizon size.
That is, it looks like a portion of the Schwarzchild metric.  As
we scale up the black hole mass we have to take into account the
presence of black holes in other horizon volumes.  More
importantly, we need to take cognizance of the fact that our
statistical arguments for the most likely state apply to all
backward light cones, including those that overlap partially with
each other.    Geometrically, it is inconsistent to insist on a
configuration, which is simultaneously spherically symmetric
around the centers of two overlapping light cones. For the
purposes of the present paper we will need to assume only two
properties of the "scaled up black holes" .  The first is that the
energy/entropy/size scalings of small black hole configurations
can be extrapolated to the regime where the black hole and
particle horizons have comparable scale.  The second is that the
geometrical features of the scaled up black holes remain
independent of their internal state.  The horizon problem is solved by
the claim that geometry is a ``thermodynamic'' property of the generic
state of a large black hole.

Indeed, we suspect that there is {\it no} consistent geometrical
description of the microstructure of the dense black hole gas.
Only its coarse grained FRW geometry makes sense.   Geometrical
features have operational meaning only when the system has
particle (or extended object) degrees of freedom which propagate
locally in the geometry.   Starting from a picture of small black
holes in each horizon volume we see that the space available for
such propagation gets smaller as we scale up the black hole mass.
In the limit of a dense black hole gas, a would be particle
emitted from one black hole is immediately absorbed by another.
There would seem to be no localized excitations of this fluid,
which is perhaps another way of understanding why it solves the
horizon problem.

The reader who absorbs and believes these arguments will have a
hard time making them consistent with our later use of field
theory to calculate fluctuations in the dense black hole fluid. We
are in complete sympathy with such skepticism, and believe that it
may explain why our calculation, although it obtains the right
power law, is not consistent with observation.  We will show that
the causality bounds of field theory (no correlation between
events that have not had time to communicate through light signals) 
prevent us from
applying our spectrum calculation to the scales at which a
Harrison-Zeldovitch spectrum has been observed in the universe.
We will emphasize that our calculation depends only on scaling
properties of the $p=\rho$ FRW metric.  We hope that these scaling
properties may survive a proper treatment of fluctuations in the
$p=\rho$ fluid, but that the field theoretic causality bounds will
be circumvented.

To summarize, if we consider a set of disjoint backward light
cones ending on our time slice, then, to the extent that the
horizon is large so that the spectrum of allowed black holes is
strongly peaked at the maximal size, observers in each of these
light cones will see the same geometry with high probability. Note
that all of the dominant fluctuations away from this most probable
situation will consist of (collections of) {\it smaller} black
holes.

Now let us argue that such a collection of maximal black holes
behaves like a $p=\rho$ fluid.   We will do this in arbitrary
dimension, and write the equations in Planck units appropriate to
the dimension in which we are working.  Let $n$ be the number
density of black holes in comoving coordinates and $M$ be their
mass. The energy density is \eqn{enden}{\rho = M n} while the
entropy density is

\eqn{entden}{\sigma = M^{{d-2 \over d-3}} n}

The black holes are all of the scale of the horizon:

\eqn{dist}{R_S \sim M^{1\over d-3} \sim H_0^{-1}}

Since there is one such black hole in each horizon volume, we have
$n \sim R_S^{-(d-1)}$

Thus

\eqn{endenb}{\rho \sim R_S^{-2}}

\eqn{entdenb}{\sigma \sim R_S^{-1} \sim \sqrt{\rho}}

This is the energy/entropy relation of a $p=\rho$ fluid.  To
really claim that we have such an equation of state we must show
that the postulated configuration of  horizon filling black holes
is stable over a reasonably long period. To see what this entails
note that with the $p=\rho$ equation of state, when the universe
expands by a scale factor $A$, then $\sigma \sim A^{-(d-1)} \sim
\rho^{1/2}$.  Using the black hole formulae, this implies that $n
\sim A^{- (d-1)^2}$.  If the black hole number were conserved we
would have $n \sim A^{-(d-1)}$ so $A^{(d-1)(d-2)}$ black holes
must merge to form a single horizon filling black hole as the
universe expands, if the black hole fluid is to give $p=\rho$.

We believe that the dense black hole fluid is classically stable.
That is, as black holes come into each other's horizon volume they
are only a Schwarzchild radius apart and will merge in less than
an e-folding time of the universe.  However, we should remember
that our claim that the state in each light cone was a maximal
size black hole is only statistical.  There is some probability
that it will be a collection of smaller black holes.  If there is
a region of contiguous horizon volumes filled with sufficiently
small black holes then they will not merge as the horizon expands.
In such a region the black holes will behave like a $p=0$ gas.
Thus, the initial state of the universe is a dense black hole
fluid with a few dilute areas where the equation of state is
$p=0$.  Since these small black holes will evaporate quickly, the
equation of state in these regions quickly becomes that of a
relativistic gas. In a coarse grained view, averaging over many
initial horizon volumes, we have a two component homogeneous
fluid.

The dilute black hole/radiation gas component has an initial
density much smaller than the dense black hole fluid component,
because the probability of finding a large enough dilute region to
prevent mergers is small.  This is a consequence of the fact that
in each horizon volume, the probability for finding any given
configuration is sharply peaked at the maximal size black hole,
because of the nature of the black hole entropy function.
Furthermore, if we choose a configuration too close to the maximum
then it will not evolve as a $p=0$ or relativistic gas.   Indeed,
imagine some collection of less than maximal black holes in a few
connected horizon volumes.  As the universe expands these black
holes will be attracted to the maximal black holes in neighboring
horizon volumes.  If the infall time is short compared to the
e-folding time in the putative $p=0$ region, then in fact the
$p=0$ equation of state is never achieved.  The matter originally
in submaximal black holes will be absorbed into the dense black
hole gas, and the empty region will cease expanding and will
become a microscopic blip in the expanding $p=\rho$ universe.
Causal processes will soon fill it with $p=\rho$ fluid.  Since the
energy density in the $p=0$ region must initially be less than
that in the $p=\rho$ region, we can only achieve the $p=0$
equation of state if a fairly large number of contiguous horizon
volumes simultaneously contain submaximal black holes.

On the contrary, if the entropy density of the fluctuation is
sufficiently small then the fluctuation region will expand
(initially according to the $p=0$ equation of state, and then like $p=
\rho /3$) more rapidly than the
surrounding $p=\rho$ fluid.   We will consider its subsequent
fate in a moment.

The fact that a finite gap from the maximal entropy is necessary
in order to produce a fluctuation which can grow, means that the
probability for such fluctuations is exponentially small in the
region of validity of the semiclassical entropy formula.  Indeed,
if the entropy of the maximal black hole is $S_{max}$ then we
expect that the probability for a growing bubble will behave like

\eqn{probubb}{P_0 = e^{- N S_{max}}}

The quantity $N$ is the number of contiguous
horizon volumes which must simultaneously have submaximal black
holes. To do a proper calculation one would have to test many
multi black hole configurations, to find the one which maximized
the entropy among all those that succeeded in breaking away from
the $p=\rho$ fluid. Furthermore, it is clear that fluctuations
produced closer to the initial time slice are exponentially more
probable than those produced later.  Thus, the only important
dilute regions are those produced at the initial instant. Indeed,
it is likely that the correct calculation of $P_0$ cannot be done
within the semiclassical approximation, since the semiclassical
entropy formulae suggest that the probability is maximized at the
point where the formulae break down.   However, since the
calculation always involves the joint probability for a number of
low probability events, the result is guaranteed to be small.
Thus, we will not attempt a calculation of $P_0$, but will believe
the strong suggestion of the semiclassical approximation, that it
is very small.

The considerations that determine $P_0$ and the dominant
configuration which breaks away from the $p=\rho$ expansion are
translation invariant, and the local probability distribution is
strongly peaked around the maximum.  Thus we expect the
distribution of dilute black hole spheres in the dense fluid
background to be uniform, with small fluctuations.   To get some
idea of how the fluctuations behave locally, consider a toy model
in which the dominant breakway configuration is a distribution of
$K$ black holes of mass $m < M_{max}$, and consider fluctuations
$m \rightarrow m - \delta m$.  $\delta m$  is presumed to be
positive because any mass larger than $m$ would be reabsorbed
into the $p=\rho$ fluid. For simplicity we will write formulae in
$4$ spacetime dimensions.   The (unnormalized) probability
distribution for $\delta m$ is

\eqn{probdeltam}{P(\delta m) \sim e^{- 8\pi K m\  \delta m }}.

If $m$ is large in Planck units, the fluctuations are of order
$1/ 8\pi K m$; small but not exponentially small.

It is clear then that the inhomogeneous fluctuations in the
distribution of $p=0$ spheres are suppressed by the same
mechanism that makes the density of such spheres small.   By the
same token, it is difficult to perform a reliable calculation of
the precise amplitude of the fluctuations.  It is determined by
Planck scale physics.  We will come back to discuss the spatial
variation of the correlation function of these fluctuations after
we have established that it determines our model's prediction for
the fluctuations in the cosmic microwave background and for
perturbations that give rise to galaxies and other large scale
structures.

The above discussion suggests that our investigation must begin at
a time when all the dimensions of the universe are in play. This
is indeed true, but we will see that higher dimensional physics
has only a small effect on our results.  First of all we must
state our hypothesis that the M-theory vacuum that is relevant for
the universe is isolated and that the moduli are frozen at a very
high scale.   This sort of scenario has long been advocated by M.
Dine \cite{dine} and has many attractions.  We adopt it both
because the alternative is more complicated (and we have not
worked it out) and because it is implied by the hypothesis of
cosmological SUSY breaking\cite{tbfolly}. If we use $M_d$ as a
symbol for the higher dimensional Planck mass, our hypothesis is
that the moduli are fixed at a SUSY minimum of a potential whose
typical scale is at least of order $M_d^6 / M_P^2$ (here, because
of the obvious ambiguity, we relax our convention that any Planck
mass is unity).

The most obvious phenomenological reason to hypothesize compact
dimensions larger than $M_d^{-1}$ is so that one can implement
Witten's explanation \cite{witstrong} of the discrepancy between
the Planck mass and the unification scale\footnote{We can, with a
bit of latitude, also add the scale of the dimension 5 operator
responsible for neutrino masses, to the evidence for a scale
around $10^{15} - 10^{16}$ GeV.}. In the Horava-Witten scenario
\cite{horwit}, where the internal manifold is highly anisotropic,
this implies a maximal Kaluza-Klein radius of order $70$ times
$M_d^{-1}$.   In hypothetical scenarios in which the standard
model arises from a codimension $4$ singularity in a $G_2$
manifold \cite{cveticetal} the coefficient $70$ would be replaced
by a smaller number. Of course, in recent years we have seen a
large number of models with much larger internal dimensions.   We
will see below that such models are problematic in our framework.

Let us thus begin our discussion of the $p=\rho$ phase in $11$
dimensions (we choose $11$ rather than $10$ for concreteness).
The mass of a black hole constituent of the $p=\rho$ fluid must
be taken larger than $M_d$ in order to justify our use of
classical general relativity in its description.   It grows like
$a^{10}$ as the universe expands. Almost immediately, its
Schwarzchild radius becomes as large as the largest $KK$
radius.   The considerations of Gregory and LaFlamme
\cite{greglaf} then show us that the entropy of eleven
dimensional black holes is swamped by that of black branes
wrapped around the compact manifold.  It is eminently plausible
(and truly the only coherent hypothesis we can make about this
highly nonclassical era) that there is a transition to a four
dimensional $p=\rho$ fluid. It no longer makes sense to talk
about internal motions on the compact manifold, since they are
(everywhere in four dimensional spacetime) hidden behind the
horizon of the wrapped black branes\footnote{The internal
dimensions will reappear at lower energies, inside growing $p=0$
regions.  In these regions we can perform scattering experiments
at the KK energy scale and produce KK excitations.}.   Within a
few instants, then, the universe becomes a dense fluid of four
dimensional black holes.  Their initial mass $M_0$ is determined
by the condition (in four dimensional Planck units)

\eqn{mzerobound}{2M_0 \simeq R_{KK}}

where we have chosen the largest dimension of the compact
manifold as a measure of $R_{KK}$ ({\it e.g.} the Horava-Witten
interval).

\subsection{Fluctuations}
\FIGURE{
            \epsfig{file=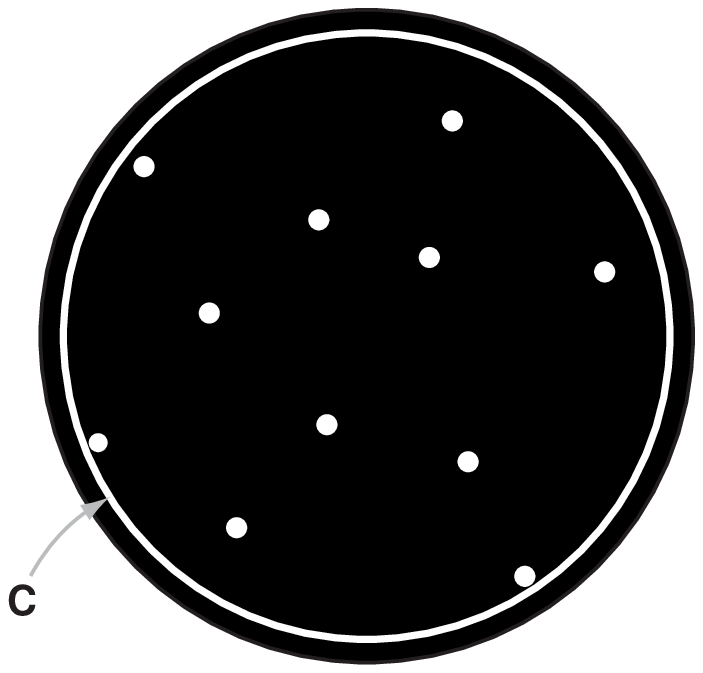}
         \caption{...}
         }

We are now ready to discuss the most important feature of any
early universe cosmology, the nature of the density fluctuation
spectrum at large scales. We will work in a synchronous
coordinate system, in which we claim that the metric has the
approximate form

\eqn{synch}{ds^2 = - dt^2 + a^2 ({\bf x}, t) d{\bf x}^2}

We begin at the initial four dimensional slice.  A two
dimensional cartoon of the initial spatial configuration is given
in Figure 1.
The outer circle in this picture represents the
fixed coordinate position of a presumed cosmological event
horizon.  Very little of what we have to say depends on this
assumption of a nonzero value of the cosmological constant.  The
black region represents the dense gas of black holes, with
equation of state $p = \rho$. The small white circles represent
the spheres of dilute fluid, while the circle $C$ is just the
boundary of the region where this finite number of spheres lie.
The finiteness of the number of dilute spheres is a consequence of
our assumption of a finite cosmological horizon.  We will mostly
discuss an approximation in which

\eqn{confac}{a({\bf x}, t) \approx a_0 (t) \chi + a_1 (t) (1 - \chi)}
where $\chi$ is the characteristic function of the region filled
with white circles. Figure 1 is very far from being accurate as
far as scale.  The density of dilute regions, $P_0$, is extremely
small.   It is important to recognize that this statement remains
true for all times.  We are about to see that most of the
physical volume of the universe becomes dominated by $p=0$
regions.

The coordinate size of the dilute spheres in the $p=\rho$
background is fixed.  Since these regions are radiation dominated
throughout most of their history, their physical size will grow like $t^{1/2}$.
Thus we have to use a much finer coordinate grid
in the dilute regions.

In any number of dimensions, in our gauge, the ratio of
the volumes of $p={1\over 3}\rho$ and $p=\rho$ fluids at any time $t$ is

\eqn{volrat}{V_{1\over 3} = P_0\ t^{1\over 2}\  V_1}
where $t=1$ is the initial time and $V_w$ is the physical volume
of regions of $p = w \rho$ fluid.   Thus, in Figure 1, the ratio
of volumes of white circles to the entire black region is $P_0
t_d^{1\over 2}$ where $t_d$ is the time between the initial instant and the
Gregory-LaFlamme (GF)transition.  $t_d$ is also, according to our
general discussion of the $p=\rho$ fluid and its black hole
interpretation, the ratio between the eleven dimensional black
hole Schwarzchild radii at the GF transition and the initial
instant.

If we assume an initial Schwarzchild radius close to $M_d^{-1}$,
then $t_d $ is approximately $R_{KK}$ in $M_d^{-1}$ units.  Thus,
for compactifications of the type we have discussed $t_d$ is at
most of order $100$. Since we have no reliable estimate of $P_0$,
we might as well define

\eqn{defpbar}{\bar{P}_0 \equiv P_0 t_d^{1\over 2}}

\def\pbar{\bar{P}_0}

Now let time evolve in four dimensions.   The ratio between the
physical volumes is $\pbar t^{1\over 2}$, while the mass of a black hole in
the dense black hole fluid is $M_0 t$.  At the crossover time
$t_+$, when $p=\rho /3$ regions begin to dominate the volume of the
universe, the mass of black holes in the $p = \rho $ regions is
$M_0 / \pbar^2 $.

We can draw a new cartoon of
the situation by rescaling the size of the white circles by the
ratio of volumes of the two fluids.  We obtain Figure 2.

\FIGURE{
            \epsfig{file=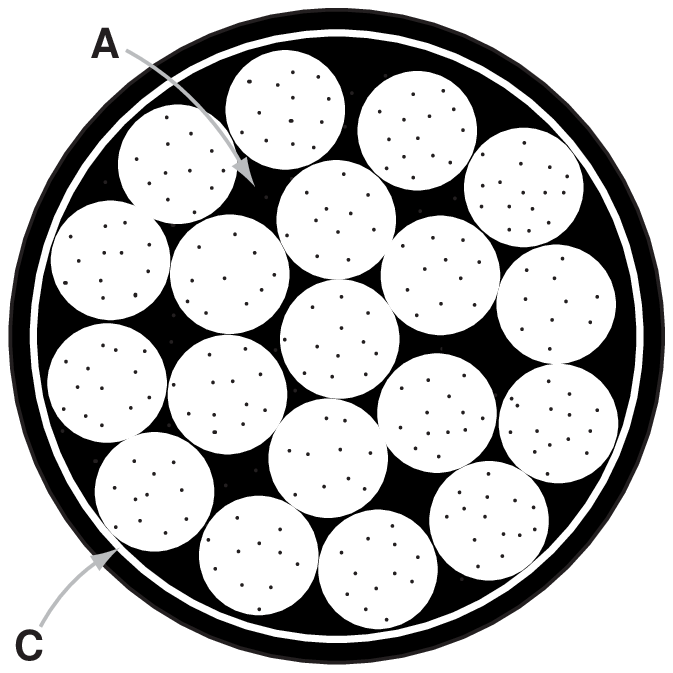}
         \caption{...}
         }
Actually, this figure also depicts the result of another physical
process, which takes place as the $p=0$ volume begins to dominate.
There are sheets of dense black hole fluid that are trapped
between expanding $p=0$ spheres.  Recall that the dense black hole
fluid retains its equation of state by continual attraction and
merger between its constituents.  Obviously, the black holes in
the sheets will not find any partners to merge with in the
directions in which they encounter spheres of dilute black hole
gas.  The expanding spheres of dilute gas will leave interstices
when they meet ({\it e.g.} the region marked A in Figure 2). The
dense black hole fluid in these interstices can continue to merge
and form larger and larger black holes.   This interstitial fluid
will attract the black holes in the sheets. We believe that all
the $p=\rho$ fluid will accumulate in the interstices.   The final
situation (and this is what we depict in Figure 2.) will be a
continuous region of dilute gas with spherical $p=\rho$ regions
embedded in it.  Each of these regions will eventually form a
single large black hole, with mass of order $M \equiv M_0/
\pbar^2$.   The entire universe (inside $C$) will now be filled
with a two component fluid. One component consists of the
radiation produced in the decay of small black holes and the other
of huge black holes with mass $M_0 /\pbar^2$. The energy density
is quickly dominated by the nonrelativistic gas of large black
holes.

  At this point
Hawking evaporation of the large black holes comes into play. We
can estimate the evaporation time of the large black holes, and
the temperature to which they reheat the universe, by writing
coupled equations for the black hole and radiation energy
densities

\eqn{coupla}{{d\rho_{BH} \over dt} = - 3 H \rho_{BH} - {1 \over M^3} \rho_{BH}}

\eqn{couplb}{{d\rho_{\gamma} \over dt} = - 4 H \rho_{\gamma} + {1 
\over M^3} \rho_{BH}}

\eqn{couplc}{{dM\over dt} = - {1\over M^2}}

\eqn{coupld}{H = \sqrt{\rho_{BH} + \rho_{\gamma}}}

We have omitted all numerical constants from these equations since
we are aiming only for order of magnitude results. Now notice
that the rescalings

\eqn{rescala}{ M \rightarrow b M}

\eqn{rescalb}{ t \rightarrow b^3 t}

\eqn{rescalc}{ \rho_{BH,\gamma} \rightarrow b^{-6} \rho_{BH,\gamma}}
leave these equations invariant.  This enables us to see how the
final photon energy density, and thus the reheat temperature,
depends on the initial black hole mass $M_0 / \pbar^2 \equiv M$.

We obtain

\eqn{trh}{T_{RH} \sim (M_0 / \pbar^2 )^{-3/2}}
This analysis also recovers the conventional estimate of the
black hole evaporation time as $\sim M^3$.

A firm bound on the reheat temperature comes from nucleosynthesis:

\eqn{trhbound}{T_{RH} > 1 MeV = 10^{-22} M_P}
One may also contemplate a stronger bound from the requirement of
generating a baryon asymmetry.  However, Affleck-Dine\cite{ad}
(AD) baryogenesis seems to work at temperatures as low as an
MeV.  Furthermore, we will see below that our scenario presents
us with a novel mechanism for baryogenesis , whose efficiency is
difficult to estimate.  Thus we will stick with the
nucleosynthesis bound.

Any semiclassical estimate of $\pbar$ based on assuming the
semiclassical entropy formula with black hole masses or
Schwarzchild radii at least an order of magnitude above the
relevant Planck scale gives a value of $\pbar$ that is completely
incompatible with the nucleosynthesis bound.   We have noted that
the semiclassical formulae themselves suggest that $\pbar$ is
determined by Planck scale physics and we don't know how to
estimate it.  The above calculation is phenomenological testimony
to the same fact.

In passing we note that the picture we have presented is
incompatible with the idea of a very large KK radius.   Whatever
$\pbar$ is, it is certainly less than one, while $M_0$ is of
order the largest KK radius.   Thus, we find that $R_{KK} < 10^{-
18} cm$. Since we believe that $\pbar$ is probably quite small,
this upper bound is probably a gross overestimate.  Holographic
cosmology does not seem to be compatible with very large KK radii.

On a more speculative note, let us consider the fact that in
Figure 2, the region that will be identified with conventional
cosmology is the interior of $C$.This sphere is surrounded by
$p=\rho$ fluid that we can eventually think of as being "squeezed
up against the horizon".  It is tempting to identify this with the
degrees of freedom on the horizon\footnote{Perhaps making contact
with the description of horizon entropy (of black holes) as a
$p=\rho$ fluid presented in \cite{tHpz}.}   Indeed, the
cosmological horizon is a holographic screen for the entire
asymptotically deSitter space time.  We use the gauge freedom
available in the formalism discussed in \cite{bf} to project some
of the degrees of freedom on local screens where more or less
conventional low energy physics can be used to describe them. The
rest of the states are associated with the horizon.  Our picture
suggests that these horizon states are remnants of a primordial
$p=\rho$ fluid.

Indeed, the part of the $p=\rho$ gas outside the circle $C$ in
Figure 1 interacts only very weakly with the local degrees of
freedom that we describe in conventional cosmological language.
Its degrees of freedom might as well be associated with the
horizon. In this context it is worth remembering that if there is
truly a cosmological constant of the order of magnitude indicated
by measurements of distant supernovae\cite{supernova}, then the
Bekenstein-Hawking entropy of the horizon is of order $10^{25}$
larger than the rest of the entropy of the visible universe (even
if we include hypothetical $10^6$ solar mass black holes in each
galaxy).

We have thus explored how our holographic cosmology produces an
approximately homogeneous isotropic radiation dominated universe
at temperatures of order $1$ MeV or above.  We are now ready to
calculate the form of fluctuations around the homogeneous state.
The primordial fluctuations in matter density are those of the gas
of large black holes caught in the interstices between $p=\rho /3$
regions.  We have sketched the way in which the distribution of
these black holes is determined by causal classical processes from
the distribution of $p=\rho /3$ spheres in the $p=\rho$ background at
the initial time.  The latter fluctuations $\delta  h({\bf x}, t)$
should be thought of as fluctuations in the synchronous gauge
metric.  We have argued that they
  are likely to be small,
though we have been unable to give a precise estimate of their
magnitude.  The result of the small size of $\delta h$ is that
$\delta \rho$ is, to a good approximation, a linear functional of
it:

\eqn{denrel}{\delta \rho ({\bf x}, t) = \int d^3 y ds f({\bf x} -
{\bf y}, t,s) \delta h ({\bf y}, s)}

$h$ is the trace part of the fluctuation in the synchronous gauge
metric $\delta h^i_j$ with indices raised by the background
$p=\rho$ metric.  The transverse metric fluctuations represent
gravitational waves.  By causality of the classical physics that
determines it, the function $f$ has a correlation length no larger
than the coordinate horizon size at the end of the transition
between $p=\rho$ and $p=\rho /3$ dominance. In fact, we believe that its
Fourier transform goes to a constant at low comoving wavenumbers
even when these waves have wavelength much smaller than this late
time horizon.   Crudely speaking, the position of a $p=\rho$
regions trapped in the interstices of the radiation dominated
phase, is completely determined by the positions of a few near
neighbor dilute regions at very early times.  Thus $f$ should vary
on the scale of the horizon at these very early times. Only
dramatic motions of the bubbles of dilute fluid in the dense black
hole background could give rise to correlations in $f$ over scales
as large as the horizon at the phase transition.  We do not see
any reason why such motions occur. If this is correct, the
complicated dynamics that fixes the large black hole positions in
terms of the distribution of $p=0$ spheres, can affect the
magnitude but not the shape of the fluctuation spectrum at low
wave number.

To calculate the shape of the spectrum we recall that for
observational purposes we are interested only in fluctuations of
very long wavelength. Furthermore, we are working in a regime
where the energy density is small compared to the Planck scale  At
very large wavelengths it is perhaps plausible\footnote{We remind
the reader that skepticism on this point is eminently justified.}
that the correlations are well described by quantum field theory.
Moreover, the dominant long distance correlations in field theory
come from single particle exchange of massless
particles\footnote{Even if part of the infrared dynamics is a
nontrivial conformal field theory, unitarity bounds guarantee that
this statement remains true.}.

Inflationary cosmology has taught us how such quantum
fluctuations are converted into classical fluctuations once they
enter the horizon.  The wave function is a superposition of
amplitudes for different classical distributions $\delta \rho_0
({\bf x})$.   Once a fluctuation is smaller than the horizon size
it causes distortions in the background geometry, which in turn
influence other degrees of freedom propagating in it.  Since
there are a large number of such degrees of freedom, phase
correlations between different pieces of the wave function become
irrelevant to the future evolution of the system. The different
parts of the wave function that produce macroscopically different
classical geometries "decohere" and quantum mechanics gives rise
to an inhomogeneous universe whose inhomogeneities can only be
predicted statistically.   The basic mechanism is in no way tied
to inflation, and is applicable to our model as well.

These considerations lead us to a Gaussian distribution for the
fluctuations.  The form of the two point function is

\eqn{fluct}{<\delta \rho (k,T) \delta\rho (-k,T)> =
\int ds ds^{\prime} f(k, T, s) f(-k, T, s^{\prime}) G(s,s^{\prime},k).}

$G$ is the (Fourier transform of the) expectation value in some
initial state of the anticommutator of the fluctuation operators
$\delta h (x,s)$ and $\delta h(x^{\prime}, s^{\prime} )$. In
keeping with our argument that the long distance fluctuations come
from free field theory, we assume that the initial state is
a Gaussian functional of $h$, and that its evolution is determined
by the free Schrodinger equation for fluctuations about the
$p=\rho$ FRW metric.

The Lichnerovitz  equation for metric fluctuations is

\eqn{lich}{D^2 h_{\mu\nu} + D_{\mu}D_{\nu} h - D^{\lambda} D_{\mu}
h_{\nu\lambda} - D^{\lambda} D_{\nu} h_{\mu\lambda} , } where all
contractions and raising of indices is performed using the
background metric, and $D_{\mu}$ is the Christoffel connection for
the background. Under a Weyl transformation $g \rightarrow
\Omega^2 g$ with constant conformal factor $\Omega$, this equation
scales by $\Omega^{-2}$, while the fluctuations $h_{\mu}^{\nu}$
are invariant.

Like all FRW universes with power law scale factors, the $p=\rho$
background has a conformal isometry.  If $t\rightarrow \Omega t$
and $x^i \rightarrow \Omega^{2/3} x^i$, the metric rescales by
$\Omega^2$.  Combined with the Weyl transformation properties of
the previous paragraph, this implies that the Schrodinger equation
for fluctuation modes of wave number $k$ has the schematic form
(it is easiest to prove this in conformal coordinates and then
transform to synchronous gauge):

\eqn{schrod}{i \partial_t \Psi =  [{\Pi^2 \over a^3} + k^2 a h^2]
\Psi .}

We have omitted indices because all that interests us is the
scaling property of the equation.   The reader should note however
that if $h$ were a scalar, this would just be the Schrodinger
equation for minimally coupled scalars.  It is well known that the
gravitational wave degrees of freedom in a general FRW background
satisfy the same equations as minimally coupled scalars.  We also
note, that although we are working in three spatial dimensions,
the crucial scaling property is valid in all dimensions.  In four
dimensions, we see the scaling by
multiplying up by $a^3 \sim t$. Then the only dependence of
the Schrodinger equation on $k$ is through the scaling variable
$|k| t^{2/3}$.  The same will be true for the expectation values
of interest to us.   

Now let us combine this scaling property with the $k$ independence
of the function $f$ relating fluctuations in the density of large
black holes to fluctuations of the metric away from the $p=\rho$
FRW geometry to obtain:

\eqn{fluct}{<{\delta\rho\over\rho} (k,T) {\delta\rho\over\rho} (-
k,T)> = \bar{f}^2 \int ds ds^{\prime}  E(s k^{3/2}, s^{\prime}
k^{3/2}) ,} where $E$ is the expectation value of the
anticommutator and $\bar{f}$ in this equation is the constant
limit of the Fourier transform of $f(x,T,s)$ for small $k$. The
range of integration goes from very small times to $T$. Rescaling
the integration variables we find

\eqn{fluct}{<{\delta\rho\over\rho} (k,T) {\delta\rho\over\rho} (-
k,T)> = {\bar{f}^2 \over k^3} \int_{s_0 k^{3/2}}^{T k^{3/2}}  du
du^{\prime} E(u, u^{\prime} ) .} 

\noindent
The dependence on the lower
endpoint drops out for the small $k$ values of interest.
Furthermore, the expectation value for each $k$ satisfies the
massive Klein Gordon equation in one dimension (time) and so, as
long as $k > T^{- 2/3}$ (wavelengths smaller than the horizon scale
at the end of the $p=\rho$ era ) there is no $k$ dependence from
the upper end point.  We have thus derived the Harrison Zeldovitch
spectrum for a range of wave lengths smaller than the horizon size
at the end of the $p=\rho$ era.  This restriction is a consequence
of field theoretic causality.   The position space expectation
values vanish for points separated by more than the horizon,
because in field theory there cannot be correlations between
events that are at spacelike separation.

Observations suggest the existence of a Harrison-Zeldovitch
spectrum of primordial fluctuations for a range of scales ranging
from that of the current horizon radius (COBE) down to the scale
of galaxies.  We must ask whether our model predicts such a
spectrum over this range of scales.   The current horizon radius
in Planck units is about $1
0^{61}$.  At nucleosynthesis, it was
smaller by a factor of $10^{-10}$.   Let us assume (since this
turns out to be the best case) that the reheat temperature $T_{RH}
\sim M^{-3/2}$ is indeed of order $1$ MeV $\sim 10^{-22}$ in
Planck units.  During the period between the end of $p=\rho$ and
nucleosynthesis, the universe is dominated by large black holes
and the universe grows by a factor of $t^{2/3}\sim M^2$, where $t$ is the
black hole evaporation time.  Thus, at the end of the
$p=\rho$ era, the current horizon volume had physical radius
$10^{51 - 88/3} \sim 10^{21}$.  On the other hand, the physical
size of the $p=\rho$ horizon at this time is of order $M \sim
10^{44/3}$, so there is a discrepancy of 6-7 orders of magnitude.
Even galaxy scales were $10-100$ times larger than the particle
horizon at the end of the dense black hole fluid era.

It is tempting to imagine that the mistake in this analysis is simply
that we have to translate comoving wave numbers in the $p=\rho$ gas into
{\it physical distances} in the dilute regions.  After all, what we are
really interested in are the fluctuations of the relative positions of
the large black holes.  However, it turns out that this does not change
the numbers very much.   In the comoving $p=\rho$ distance of order the
horizon scale, $M^{2/3}$ , (this is the distance between two typical black
holes that can be correlated by the causal fluctuations we have
calculated) one encounters $\pbar^{1/3}$ dilute regions.  Each of these
regions has linear physical size $M^{1/2} N^{1/3}$ times the horizon
scale at the beginning of the four dimensional era.  $N$ is the number
of contiguous horizon volumes at the beginning of the $p=\rho$ era that
are included in one dilute region.  Remember that $N$ has to be larger
than $1$ in order for the dilute regions to grow, but it cannot be huge.
In particular, $\pbar \sim v^N$ with $v <1$.  Since $\pbar \sim M^{- 1/2}
\sim 10^{-7}$,  $N^{1/3}$ cannot be very much larger than $1$.

Thus, we find a typical distance between correlated black
holes of order $M N^{1/3}$. Apart from the (order $1$) factor $N^{1/3}$ 
this is the same as the estimate we made above, measuring distances in
the $p=\rho$ background.  The reason they are the same is that we have
chosen the transition time to be the point where volumes in the two
fluids are equal.   Thus, our discrepancy cannot be attributed to
confusion of the scales of distance in the two geometries.

We have tried to resolve this discrepancy by using conventional
astroparticle methods, but to no avail.  For example, one could
try to abandon the idea that the decay of large black holes gives
rise to the Hot Big Bang.  Instead, introduce a particle species
that produces the usual radiation in its decay, and allow the
large black holes to have a much lower reheat temperature.   This
allows us to extend the length of the $p = \rho$ era, and make the
current horizon fit inside the $p=\rho$ horizon.   The problem is
that it does not make sense to talk about particles until the end
of the $p=\rho$ era.   Thus, we see no reasonable way to insist
that the energy density in large black holes be many orders of
magnitude smaller than that in the decaying particles.   Thus, the
universe will not in fact become radiation dominated.   It will
remain dominated by black holes until they decay.

Another possible scenario, a late period of inflation, which would
stretch out the scale of perturbations generated during the
$p=\rho$ era, until they were at least as large as the current
horizon, fails for similar reasons. It is not implausible that if
the universe contains the right kind of weakly coupled scalar
fields, that some of these fields are not at their minima at the
end of the $p=\rho$ phase, and undergo a period of friction
dominated motion. In order to have inflation, the potential energy
of these scalars must dominate the energy of the gas of large
black holes. Inflation of the scale factor by $10^6 - 10^7$
dilutes the density of large black holes by a factor of $10^{18} -
10^{21}$. The energy density is dominated by the inflaton field,
which eventually decays.

There are now two possibilities.  Either the inflaton creates the
hot Big Bang, and the black holes are irrelevant for the
explanation of density fluctuations, or the inflaton decays early
enough so that the black holes can dominate the universe once
again.   It makes sense to discuss ordinary scalar fields and
inflation only after the end of the $p=\rho$ era, when local field
theory is a good approximation.   This means that the inflationary
energy density cannot be larger than $M^{-2}$ .  If $M$ is too
large, the inflationary energy density will be low and it will be
difficult for the standard inflationary mechanism to generate
density fluctuations that are large enough to be consistent with
observation.

Early inflaton decay, on the other hand is completely
inconsistent.  Let $\alpha$ be the factor by which the length
scales are stretched by inflation.   Then, assuming inflatons
decay before black holes, and that the black hole reheat
temperature is $1 MeV$, we must have \eqn{stretch}{\alpha M^3 \sim
10^{51} \rightarrow \alpha = 10^7} in order for the current
horizon to be within the $p=\rho$ horizon.  But in order for the
black holes to be the dominant energy density at $1$ MeV we must
have

\eqn{infrh}{{\rho_{BH} \over \rho_I} = \alpha^{-3} {T_I \over
10^{-22}} > 1.} This implies that the inflaton reheat temperature
$T_I> 10^{-1}$ in Planck units. This is incompatible with the
requirement that inflation not begin until after the end of the
dense black hole fluid phase. The latter requirement implies that
the energy density during inflation satisfies $\rho_I < M^{-2} $,
and in our model $M > 10^3$, because of dimensional reduction.
Certainly $\rho_I> T_I^4$ and typically $\rho_I >> T_I^4$ for a
weakly coupled inflaton.  One could attempt to construct a model
in which the dense black hole gas phase ended while the universe
was still eleven dimensional, and was followed by a period of
inflation, but it is easy to see that even in this context we
cannot save the idea that the observed density fluctuations are
generated during the $p=\rho$ phase.

The only inflationary escape from our problem is to postulate that
the dense black hole fluid phase of the universe terminates rather
early and that all observable features of the universe are
produced by a subsequent inflationary phase.   In order to
reproduce the observed amplitude of fluctuations without excessive
fine tuning, the inflationary energy density must be at least
$\rho_I > 10^{-14}$ in Planck units\footnote{We are well aware
that the inflationary literature is replete with models that
reproduce the data with lower energy scales.  Our attitude towards
field theoretic fine tuning is somewhat more austere than that of
many of our inflationary colleagues.}.  This is consistent with
the bound $\rho_I < 1/M^2$ with $M > 10^3$.   Nonetheless, we feel
that it is unlikely that a fundamental calculation of $M$ will
give a large enough value to be consistent with such an
inflationary model.  Our semiclassical estimates were not even
consistent with the much weaker requirement $M > 10^{44/3}$. If
this is true, we have a very interesting situation:  {\it We
believe that the $p=\rho$ fluid is a very robust and plausible
model for the initial state of the universe.   On the other hand
we are finding it difficult to reconcile that statement with the
observed density fluctuations.  We must either find an explanation
for the discrepancy in length scales in our model, or convince
ourselves that a very short $p=\rho$ phase, compatible with
reasonable inflationary models of the fluctuations is plausible.}
We will discuss this question further in the conclusions.

  Before
proceeding to a tour through the other famous problems of
cosmology, we want to emphasize again the reason for our optimism
about the robustness of the initial conditions from which we
began.  At the heart of our argument are the holographic principle
and the UV/IR connection.  The first principle allows us to state
with confidence that the $p=\rho$ fluid is a generic initial
condition for the universe.  It saturates the FSB entropy bound
and it is the stiffest equation of state compatible with either
this bound, or causality. As a consequence it dominates physics at
high energy density. Of course, energy density is only defined
once we are in the semiclassical regime where at least coarse
grained geometry makes sense.  From this point of view as well one
is led to the $p=\rho$ gas as the generic state compatible with
the entropy bound within a given particle horizon.

Although we have called it semiclassical, the $p=\rho$ regime is
not one in which low energy field theory is applicable to most
calculations.  If we had to understand the detailed quantum
mechanics in this regime, we would be at an impasse.  The UV/IR
connection comes to our aid here.  It assures us that we can
understand the high energy spectrum of the quantum theory of
gravity in terms of black hole physics\footnote{This version of
the UV/IR connection has been advocated in \cite{bfbhetc}}.  As
long as we are willing to ask sufficiently inclusive questions
(inclusive cross sections, decay rates {\it etc.} rather than
detailed exclusive amplitudes) semiclassical GR and the Hawking
radiation formulae give us an adequate picture of the physics.
Apart from one point, this is all we have really needed for
cosmological purposes.   The one piece of quantum mechanics we
have used is the assumption that quantum correlations at extremely
large spacelike separation in the $p=\rho$ fluid are dominated by
single graviton exchange.  We have argued that this use of field
theory was justified, since the local curvature is small, the
exchanged momentum is extremely small and we are calculating
correlations within a fixed event horizon.  Still, this
calculation has led to an impasse.  It is likely that the
resolution of this problem will require us to understand the
quantum mechanics of the $p=\rho$ regime at a much more
fundamental level.

\subsection{The monopole problem}

We now turn to another conundrum of Big Bang cosmology, the relic
density of magnetic monopoles.  At first sight this will appear
to require a major revision of our estimates in the previous
section, but we assure the reader that when the dust settles,
everything is as before.   Monopoles are only defined once we
enter into the four dimensional regime.   In higher dimensions
the $U(1)$ of electromagnetism is likely to be unified into a
nonabelian gauge group. At any rate, the system goes through a
complicated Gregory-LaFlamme transformation just before the four
dimensional regime, and it is unlikely that we will be able to
make more than probabilistic statements about monopoles.
Fortunately, the initial probability distribution is highly
peaked.

Thus, we look at the four dimensional dense black hole fluid at a
time when the typical black hole mass is $M_0$, and ask for the
probability that the black hole is charged. For small charge,
that probability is $e^{-2\pi q^2}$.  Here $q^2  = n^2 g^2 /8\pi$
where $n$ is an integer, and $g$ is the magnetic charge, related
to the fine structure constant by $\sqrt{4\pi\alpha} g = 2\pi$.
At these high energies, we should use the grand unified value,
$\alpha \sim {1\over 25}$.  That is, $g = 2\sqrt{2}\pi$.  The
probability that a black hole has integer charge $n$ is thus
$e^{- 2\pi^2 n^2} \sim (3 \times 10^{-9})^{n^2}$. The probability
for all but the smallest charge is negligibly small.

However, the black holes continue to merge until their typical
mass is $M = M_0 / \pbar^4 $.  At this time $\pbar^{-4}$ black
holes have merged to form a single one.  At the initial time, the
typical charge in a region which will eventually form a single
black hole is the square root of the total number of charge $\pm
1$ monopoles in that region, or $\sim 10^{-5} \pbar^{-2} = 10^{-5}
(M/M_0)$. This random distribution will be biased somewhat by the
magnetic forces between charges, but causality will not allow most
of the charge to annihilate \footnote{We thank Raphael Bousso for
emphasizing and reemphasizing this point to us.}.Thus, our
estimate is high by a factor of order one. There is an even larger
electric charge ($10^{-5} \rightarrow 10^{({-5\over 16\pi^2})}$ in
the estimate above)per black hole at the end of the $p=\rho$ era.
However, most of that will be blown off as electrons or positrons
and quarks in the Hawking radiation we described in the previous
section \footnote{Gibbons\cite{gibbons} has argued that dyonic
black holes may not completely discharge themselves.   We believe
that this mechanism will only operate in the regime where electric
and magnetic charges (probably weighted by the corresponding
couplings) are of the same order of magnitude.   Thus, it is
possible that the remnants discussed in the text have comparable
electric and magnetic charges.}.  The same is not true for
magnetic charge.  The charged black holes are large, their Hawking
temperature is too low and the magnetic field at the horizon too
small, to give significant probability for emitting GUT scale
magnetic monopoles. Instead, if there is no annihilation of
magnetically charged black holes during the $p=0$ era, they will
end up as extremal black holes of very large magnetic charge.

It is thus crucial to calculate the annihilation rate of these
charged black holes.  The problem is in principle quite
complicated, because it involves black hole scattering, expansion
of the universe, and Hawking evaporation.  However, the time
scale for the latter is long and we will see that the initial
monopole density is so low that annihilation does not occur.  The
time scale for the monopoles to reach their asymptotic density is
short compared to the Hawking evaporation time. This simplifies
the problem because we do not have to worry either about the
change of mass of the monopoles or the contribution of Hawking
radiation to the cosmological expansion.   The rate equation for
the monopole number density is thus

\eqn{rate}{{dn \over dt} = - <\sigma v> n^2 - 3H(t) n}

Here $\sigma$ is the monopole annihilation cross section and $v$
the average rate of collision.

Since the monopoles are initially fairly far from extremality,
their interactions are dominated by gravitational forces.  The
cross section for annihilation will be larger than that for merger
of two monopoles of the same charge, but only by a factor of order
one.  It should be clear that even when we speak of annihilation
of oppositely charged monopoles, we mean only that their charge is
cancelled.  These are huge black holes and will make larger ones
upon merger, no matter what their charge.

The annihilation cross section is at least as large as $\pi
(4M)^2$.  This is the area of a disk with radius the Schwarzchild
radius of the combined system. Black holes can be captured into
bound states at even larger impact parameters than this.

We estimate the velocity in this expression as follows: a zero
energy solution of Newton's equation for two black holes
interacting by gravitational attraction satisfies

\eqn{zero}{\dot{r}^2 \sim M/r}

It is easy to see from this that the time to travel multiples of
the Schwarzchild radius scales like $M$.  Thus, for a dilute
black hole gas, $v$, the frequency of collison, is of order
$1/M$.  Thus $\Gamma \equiv <\sigma v> \sim 32\pi M$.

If we consider $H(t)$ a given function of time, it is easy to
solve the rate equation for $n$, by introducing the variable $x
\equiv {1\over n}$.   The equation is linear in $x$.   Its
solution involves time integrals of $H$ which are of course just
the log of the scale factor.  We write the solution as a function
of the scale factor, choosing $a=1$ at the beginning of the $p=0$
era.  If $n_0$ and $\rho_1^0$\ \footnote{We use this notation to
emphasize that this is the energy density of $p=\rho$ fluid at
the beginning of the $p=0$ era.} are the monopole number density
(which at this time is the same as the black hole number density)
and energy density at this time then

\eqn{nsoln}{n = {n_0 \over a^3 (1 + {3n_0 \Gamma \over 2
\sqrt{\rho_1^0}}[1 - {1\over \sqrt{a}}])}}

For large $a$ this approaches $n_* / a^3$ with

\eqn{nstar}{n_* = {n_0 \over (1 + {3n_0 \Gamma \over 2
\sqrt{\rho_1^0}})}}

$n_*$ is thus a relic monopole density, which will redshift like
all other nonrelativistic matter, and more slowly than radiation.
Now note that

$n_0 \sim {1\over M^3}$, $\rho_1^0 \sim {1\over M^2} $ and $\Gamma \sim M$.

The annihilation is inefficient for large $M$ and each black hole
will leave behind an extremal remnant with mass about
$10^{-5}(M/M_0)$.  This result is not unexpected.  We defined the
transition point between dense and dilute black hole fluids by
the requirement that black holes not merge efficiently.  Thus we
should not be surprised that their annihilation cross section is
small in the dilute phase.

Recall that the reheat temperature is of order $T_{RH} \sim
M^{-3/2}$ and that the number of photons produced per black hole
decay is ${M\over T_{RH}} \sim M^{5/2}$. Thus, at reheating, the
ratio of the number densities of extremal black holes and photons
is $M^{-5/2}$.   At any lower temperature (assuming no entropy
production) the ratio of monopole to photon energy density is thus

\eqn{monrat}{{\rho_M \over \rho_{\gamma}} \sim {10^{-5} M^{-3/2}\over M_0 T}
\sim 10^{-5} {T_{RH} \over M_0 T}}

In order for nucleosynthesis to proceed in a normal manner,
$T_{RH}$ cannot be below an MeV.  Thus, at the time when the
universe was ''observed" to go through matter radiation equality,
$T_{RH}/T > 10^5$. On the other hand, even in a more or less
isotropic compactification of eleven dimensions to four, $M_0$
(in Planck units) is somewhat bigger than the KK radius, in order
to have $M_{GUT} < M_P$.  If we identify the KK radius with
$(M_{GUT} = 2 \times 10^{-3})^{-1}$ we find a monopole to
radiation ratio of about $10^{-3}$ if we assume the minimal
reheat temperature.

Several things are clear about this estimate.  First, the reheat
temperature cannot be too much larger than $100$ MeV.  This poses
a challenge for baryogenesis, which we will discuss below.
Second, the many factors of order one that we have neglected
could be important.  Thirdly, and more interestingly, the
question could be sensitive to issues such as the details of the
compactification from higher dimensions, and the question of the
precise relation of the unification scale to the KK geometry. For
example, the recent indications of neutrino masses call for a
scale an order of magnitude or so below the unification scale.
Perhaps this is closer to the KK radius than the unification
scale is.

Thus, in our model, the solution of the monopole problem requires
a low reheat temperature. It is interesting to see the question
intersecting with some of the aspects of string theory one would
have thought were completely out of reach.

In conventional discussions of the monopole problem, the Parker
bound coming from the persistence of galactic magnetic fields, and
the bound from monopole catalysis of proton decay give stronger
constraints than the energy density bound we have used.   This is
not the case in our model, since the black monopoles are such
heavy objects.  The other bounds depend on the number density and
flux of monopoles, rather than their energy density.   If, to be
concrete, we take a reheat temperature of order $1$ MeV and $M_0
\sim 10^3 $ in Planck units, then our monopoles have mass $10^7
M_P$ .  For a given energy density, their number density is
smaller by a factor $\sim 10^{-10} (M_{GUT}/10^{16}$ GeV) than
corresponding GUT monopoles\footnote{It is tempting to speculate
that a large fraction of the black monopoles might have collected
in the core of the galaxy during the process of its formation, and
have something to do with the generation of the galactic magnetic
field.  We have not made any serious attempt to estimate whether
this is plausible.}.  Furthermore, unlike GUT monopoles, the
gravitational interactions of black monopoles with neutral matter
will dominate over their interaction with the galactic magnetic
field.

  Thus we believe that the Parker bound and the bound from
catalysis of proton decay (which would anyway have to be rethought
for these black monopoles) do not put further constraints on our
model. Of course, the most interesting thing about predicting the
existence of such objects is that they might one day be found.

\subsection{Baryogenesis}

The low value of the reheat temperature that we have had to invoke in order
to solve the monopole problem puts constraints on possible theories of
baryogenesis in our model.  We do not find this particularly worrisome, since
Affleck-Dine baryogenesis can operate at very low temperatures.
After the end of the $p=\rho$ phase (which, for the values of parameters
we have estimated in the previous section occurs at an energy scale 
about $10^{12}$ GeV),
conventional field theoretic analysis of the evolution of the
universe is sensible.  There is no reason to imagine that the average values
of fields with very small potentials are at their classical minima, 
nor that they have the
same value in each horizon volume.  We have not analyzed
the AD scenario in detail but we see no reason why it should not work.

We want to point out however that there is another potential
source of baryon asymmetry in our model.  There is no reason for
the Hawking decay of black holes to conserve baryon number.
Indeed, neutral black holes are expected to emit equal numbers of
baryons and antibaryons no matter how much baryon number there
was in the initial state that formed the black hole. Electrically
charged black holes will of course show a preference for emitting
baryons of the same charge. In a CP symmetric world,
this asymmetry would be cancelled by emission from black holes
with the opposite charge.

The $\theta$ parameter of QED cannot be measured by any conventional 
experiment.
However, it gives a (generally irrational) electric charge to any magnetic
monopole.   This will be reduced during Hawking evaporation to a 
value lower than
the charge on the electron, but not to zero.   Unless CP is an
exact symmetry spontaneously broken at low energies, there is no reason
to assume that $\theta_{QED}$ is small.  Thus, monopoles are an intrinsic
source of CP violation.  It is then quite obvious that the Hawking decay
of our magnetically charged black holes will produce a baryon asymmetry.
Whether this is large enough to account for all of the baryons in the 
universe is less obvious.
It seems plausible to us that one can do this calculation
using the technology available to us ({\it i.e.} without having to invoke
Planck scale physics on the decaying black hole horizon), but we have 
not yet done it.
It is an interesting project for future work.

To summarize, although we see no reason to worry about baryogenesis 
in our model,
we do not yet have a calculation.  Both the AD mechanism and CP 
violating decays of
black holes provide possible origins of the baryon asymmetry 
compatible with holographic cosmology.

\subsection{The flatness problem}

The flatness problem is usually stated in the context of
homogeneous isotropic models of the universe.  Such models are
characterized by a continuous parameter, the initial spatial
curvature in Planck units at a time when the energy density is
considered low enough for the semiclassical approximation to be
valid.  Comparing the models to data, and eschewing the
possibility of an inflationary period, one finds that one must
fine tune this initial parameter drastically, in order to agree
with observation.  Inflation solves this problem by guaranteeing
that any initial spatial curvature is rescaled to a sufficiently
small value before the epoch of classical cosmology begins.

On the other hand, homogeneous isotropic models have another
problem, the horizon problem.  Since the invention of inflation ,
cosmologists have felt that the homogeneity and isotropy of the
observed universe should be explained in terms of more generic
initial conditions, which take into account the constraints of
causality.  The fact that inflation solves both of these problems
at once has obscured the fact that the conventional statement of
the flatness problem depends on the {\it a priori} assumption of
homogeneous isotropic initial conditions that is rendered suspect
by the horizon problem.  One would like a statement of the
flatness problem that does not make this assumption, and indeed
makes no hypothesis about the global spatial topology of the
universe.  The latter condition is particularly warranted if the
universe has a nonzero cosmological constant.  In that case, there
are many solutions of the equations of motion for which the global
spatial topology is never observable.  Phenomena connected with it
are forever outside the cosmological horizon of any observer.

In this section we will show that our cosmology does explain a
preference for spatially flat cosmology within the horizon scale.
The arguments for positive and negative spatial curvature are
quite different.   Let us return to the beginning of our
considerations, where we argued that a $p=\rho$ fluid can
saturate the entropy bounds.  This argument was carried out in
flat space for a good reason.   It is simply untrue in negatively
curved space. Although any equation of state can be chosen to
saturate the holographic bound at a particular time, generic
equations of state in flat space will then violate it at earlier
times and fail to saturate at later times.  In negatively curved
space this is true even for $p=\rho$.  The crucial point is that
for large enough area, the coordinate area and volume scale the
same way. Thus, it is simply impossible for the physical area of
the holographic screen to scale like the coordinate volume of the
region it encloses and the holographic bound cannot be saturated
with a homogeneous coordinate entropy density.

This has two implications.  For fixed area of the initial backward lightcones,
the negatively curved FRW geometry has less entropy than the flat one
and so is a much less probable initial condition.  Secondly, even if one
tries to insist locally on negative curvature, one finds that generic initial
conditions will be highly inhomogeneous and the description of the universe
by an approximate FRW geometry will fail.  So there is simply no analog of
our model in which the flat spatial sections are replaced by 
negatively curved ones.

For positive curvature, the argument is even simpler.  A region of positive
spatial curvature of order the horizon scale will collapse in a time of order
the e-folding time.  Thus, if we imagine randomly sprinkling flat and positive
curvature regions among the horizon volumes on the initial surface, 
the positive
curvature regions will quickly become a sprinkling of black holes in a flat
background  {\it i.e.} they will simply fade into the dense black hole gas.

One might try to get around these arguments by postulating a
radius of spatial curvature much larger than the horizon size
during the dense black hole phase but much smaller than our
current horizon.  A negative curvature of this order would not
substantially degrade the entropy of the initial conditions, while
a positive curvature of this order would not lead to immediate
collapse. Apart from the fine tuning implicit in these initial
conditions, it would be totally at odds with the concept of
particle horizon to choose this curvature to be the same or even
the same sign in each horizon volume.   Thus, what one is really
discussing with this proposal are small fluctuations of spatial
curvature around a flat FRW background.   It is well known
\cite{tk} that fluctuations in spatial curvature are nothing else
but adiabatic density perturbations.   We have discussed our model
for the origin of density fluctuations above.  It does not give
rise to an average curvature.

Thus, we claim that holographic cosmology implies a flat FRW
universe with small curvature fluctuations/density fluctuations.
We make no statement about the global structure of this FRW
cosmology and given the observed acceleration of the universe it
is possible that no such statement would have an observational
meaning. A proper understanding of the initial conditions for
cosmology shows that there is no flatness problem to be solved,
and the observational verification that the universe is at
''critical density" should be viewed merely as a test of
Einstein's equations rather than as evidence for inflation.

\subsection{The entropy problem}

Another problem whose solution is often attributed to inflation
is the {\it entropy problem}. We must confess to never having
understood the statements of this problem in the inflation
literature.  Entropy is a measure of the genericity of a state
given certain {\it a priori} conditions. In order to be able to
discuss whether a given state has more or less entropy than
expected, one has to understand what all the states of the system
are and what generic initial conditions are.   The answer to the
first of these questions was unavailable to us until recent
developments in string theory (verification of the
Bekenstein-Hawking entropy formula by microscopic calculations,
and the UV/IR connection) enabled us to begin to form a coherent
picture of the Hilbert space of quantum gravity. The answer to
the second question is still shrouded in mystery.   We have made
a fairly concrete proposal about it in this paper and in
\cite{bf}.   Alternative discussions of this question
\cite{villindehartlehawking} rely on semiclassical field theory
calculations in a regime where their validity is questionable.

So what is the entropy problem?  One way to understand an answer to this is
to examine the proposed inflationary solution to it.   As we understand it
this consists of the explanation of how one naturally gets a Hot Big 
Bang universe
with a sufficiently high temperature for classical cosmology,
from inflationary initial conditions {\it i.e.} reheating.  To the extent that
this is the statement and resolution of the entropy problem, then it has
an analog in our model, which we have discussed at length.  The 
crucial parameter
that determines the reheating temperature depends on Planck scale 
physics and we have
been unable to calculate it.  This is unfortunate in a certain sense 
and exciting in another.
On the one hand it means that our model cannot, in its current state, 
make a falsifiable prediction
about this parameter.  On the other hand
it means that an already measured quantity can be used to test the 
complete theory of
Planck scale physics once it has been constructed.

Another statement about early universe cosmology that is 
mathematically equivalent
to at least some statements of the entropy problem in the literature 
is the following:
if one extrapolates our current horizon volume back to the
Planck energy density using noninflationary Big Bang cosmology, then 
its volume in
Planck units at that time is $10^{88}$.   This is a large
dimensionless number which seems to demand an explanation.  The holographic
principle indeed connects this number to entropy.   The observable universe
requires a Hilbert space of a minimal size for its description.  This 
is determined
either by the entropy of the microwave background or, if there are 
indeed black holes of
comparable size in the centers of all galaxies, by the entropy of 
''observed" black holes.
The holographic principle states that
this number of states is not compatible with a geometrical picture of 
size smaller than $10^{88}$.

In an FRW cosmology with vanishing cosmological constant, the number of
possible states of the system is infinite, as is the entropy of the photon gas.
In such a model the number $10^{88}$ reflects only the age of the universe and
the finite number of its infinite collection of states that we have been able
to see.  Thus, this number is only a puzzle, to the extent that we do 
not understand
the small value of the cosmological constant.   In our opinion, this 
is not a soluble problem.
The cosmological constant is a measure of the total
number of states $N$ necessary to describe the universe.  In 
conventional quantum mechanics,
the number of states of a system is not derived dynamically, but is 
given {\it a priori}.

Thus, we believe that our model provides a ''solution of the
entropy problem" {\it i.e.} the explanation of a Hot Big Bang
with the requisite temperature, that is a viable alternative to
the explanation provided by inflationary theories. The crucial
parameter $\pbar$, which determines the reheat temperature,
depends on a combination of Planck scale and GUT/Kaluza-Klein
scale physics and is at the moment uncalculable. In our opinion,
nothing could be a more interesting challenge.

\subsection{Gravitinos, axions and dark matter}

In the late stages of evaporation of the magnetic black holes,
the Hawking temperature gets quite high (of order $10^{14}$ GeV
for the parameters we used in previous sections) and there will
be copious gravitino production in any model of SUSY breaking.
This could be problematic for scenarios with gravitinos at the
weak scale, like gravity mediated models.  We have not yet done
detailed estimates of the relic gravitino density, so we do not
know if this problem is real.   One of the authors (TB) has
recently suggested\cite{tbfolly} that SUSY breaking is
cosmological in origin and that the gravitino mass is of order
$\Lambda^{1/4}  \sim 10^{-3} eV$. This makes it stable even if
R parity is violated.  On the other hand, it is relativistic and
does not upset the picture of radiation dominated nucleosynthesis.
Since it does not participate in the entropy production due to
hadronic rescattering of black hole decay products, its density
is lower than that of other relativistic species.  Another feature
of such a gravitino is that its longitudinal components have
$TeV$ scale couplings.  Thus even if there is a conserved R
parity, or similar symmetry there will be no cosmologically
stable LSP which could be a candidate for cold dark matter.

On the other hand, if, as our estimates suggest, the reheat temperature
is of order the nucleosynthesis temperature, then DFSK\cite{dfsk} axions
with a GUT scale decay constant provide both a solution of the strong 
CP problem and
an excellent dark matter candidate \cite{bdaxion}.   Indeed, such a field
would be stuck away from its minimum during cosmological evolution, until the
Hubble parameter becomes equal to its mass

\eqn{hubeqmas}{H = \sqrt{\rho}/m_P \sim m_a = \sqrt{\rho_a}/f_a}

where $m_P \sim 2 \times 10^{18}$ GeV is the reduced Planck mass.
Thus, the ratio of axion energy density to the total is $f_a^2 /m_P^2$.
In our model, the total energy density is dominated by the nonrelativistic
gas of charged black holes down to the reheat temperature.
Thus the ratio of densities remains constant. Below this
temperature the axion to radiation ratio is given by
\eqn{rhoag}{\rho_a = \rho_{\gamma} {f_a^2 \over m_P^2} {T_{RH} \over T}}
Thus, the axion and radiation densities are equal at the canonical value
$\sim 10 $ eV of the temperature if
\eqn{ax}{f_a = m_P (10^{-5} {1 MeV \over T_{RH}})^{1/2}}
For a reheat temperature of $1$ MeV this gives $f_a = 5 \times 10^{15} $ GeV.

In string theory compactifications incorporating Witten's
explanation of $M_{GUT} / M_P$, there are many gauge bundle moduli
which have the potential to couple like QCD axions with decay
constants around the unification scale.

\section{Comparison with inflationary models, and conclusions}

We have made a continuous comparison of our model's properties with those of
inflation throughout the course of the previous section,
so we will try to be brief.
We present the comparison as an itemized list:

\begin{itemize}

\item Our model is based on a real, though incomplete, theory of initial
conditions in quantum cosmology, and fully incorporates the
holographic principle.  By contrast, the genericity of inflation,
though often discussed, cannot be reliably assessed. Inflation is
attractive because it washes away many of the bothersome details
of the initial condition problem.  But it has not been established
that inflation occurs with high probability in a fundamental
theory.

\item Holographic cosmology is rather uniquely defined and has very
few moving parts. It has two parameters, which determine the
reheat temperature of the universe and the amplitude of density
fluctuations.  The first, $\pbar$ depends on a combination of
Planck scale and GUT scale physics, while the second (to which we
have not given a name) depends only on Planck scale physics. A
third parameter, $M_0$ could be calculated for a given
compactification scheme. Both of the parameters that depend on
Planck scale physics must be small, and we have given a
qualitative explanation of why this is so.

We have given a qualitative explanation of much of cosmology in
terms of this small set of parameters. It is even possible that
the baryon asymmetry can be explained in terms of the Hawking
decay of magnetic black holes, which are the central actors in our
cosmological drama.  If this is correct, then only the explanation
of dark matter would seem to require a conventional particle
physics mechanism. Given the fact that our model is constrained
(by phenomenology) to have a reheat temperature close to
nucleosynthesis temperatures, a DFSK axion with GUT scale decay
constant is an attractive candidate. By contrast, there are a
plethora of inflation models and many of them are extremely
complex. Popular models all require unnatural fine tuning of
parameters to fit the observed amplitude of density fluctuations,
and concomitantly they predict a huge number of e-foldings of
inflation. Modular inflation models\cite{bgetal} require only mild
fine tuning (to get inflation at all - density fluctuations work
perfectly if the superpotential is GUT scale)and predict a minimal
number of e-foldings, but they are not very popular. Inflation
allows for a wide variety of reheat temperatures and is compatible
with a zoo of particle physics explanations of baryon asymmetry,
dark matter and the like. Even the spectrum of density
fluctuations is not uniquely predicted and depends somewhat on the
inflationary potential.  Similarly, there can be a variety of
predictions for the amplitude of the relic gravitational wave
spectrum.

\item The existence of cosmic acceleration\cite{supernova} suggests
(though of course does not prove) that there is a cosmological
constant. This makes us uneasy about inflationary theories that
require a huge number of e-foldings. The inflationary explanation
of density fluctuations requires us to believe in the existence of
independent degrees of freedom in disjoint ''temporary horizon
volumes". This is perfectly consistent with the holographic
principle and Cosmological Complementarity\cite{bf} if the
universe has no fixed cosmological horizon (though we are less
sure of this statement for cosmologies with quintessence).   In a
universe with a cosmological constant most of these degrees of
freedom are however, in principle unobservable. Complementarity
tells us to think of them as gauge degrees of freedom.  We have
not been able to think of a sharp argument that this implies that
the quantum treatment of these models is inconsistent with a
nonvanishing cosmological constant, but the oft heard claim that
our horizon volume is a tiny part of a huge inflationary patch
seems to suggest a contradiction. Quintessence theorists will of
course have no problem coming to terms with this argument, as will
modular cosmologists\cite{bgetal}, whose theories do not predict
huge numbers of e-foldings.

\end{itemize}

Against these advantages we must place the main disadvantage that
the present model has.  It is highly constrained and predicts a
Harrison-Zeldovich fluctuation spectrum in only a small range of
scales. According to our calculations the largest scale in this
range is smaller than the current Hubble radius by a factor of
$10^6 - 10^7$.  The constraint on the range of scales comes from
the causality requirement of field theory.  Scale invariant
fluctuations are generated during the dense black hole gas phase
of the universe.  In a field theory calculation, such fluctuations
must have wavelength shorter than the horizon scale of the dense
black hole gas at the end of the $p=\rho$ era.  As noted above,
the phenomenological constraints on our model set this scale at
less than $10^{-7}$ of the present Hubble radius.

Our decision to use field theory to calculate the fluctuations was
motivated by the fact that we are interested in very long
wavelengths and that the energy density is substantially smaller
than the Planck scale.  We suspect that the discrepancy we have
found is due to the fact that field theory is not a valid
description of fluctuations in the $p=\rho$ phase.  Indeed, most
of the states of this system should be thought of as near horizon
states of black holes.

The holographic principle tells us that we can think of all the
states describing physics inside a backward light cone as being
located on the FSB area of that cone.  It is plausible that the
information about localized states in the bulk is encoded locally
on the FSB surface.  We imagine the code as consisting of a bundle
of light rays from the localized object to the FSB
screen\footnote{The choice of direction of these rays is a gauge
symmetry, which one of the authors\cite{tbfolly} has conjectured
is related to local SUSY.}.  On the other hand, a black hole which
fills the horizon volume will be projected on the entire FSB
surface (via the hedgehog map from the black hole's apparent
horizon to the FSB surface).  It seems reasonable to assume that
(since the black hole is in thermal equilibrium) a typical state
of the hole involves correlations between distant points on the
FSB surface.

Now consider two disjoint particle horizons.   Typical localized
states in these regions will be out of causal contact.  But it is
often the case that the points of closest approach of the two FSB
surfaces are in causal contact.  If our guess about the nature of
black hole states is correct, this implies that two black holes in
disjoint horizon volumes during the $p=\rho$ era, are in causal
contact.   We have not yet understood how reasoning like this
could lead to acausal (from the point of view of local field
theory) correlations over distances more than a few times the
horizon scale, but it surely implies more correlation between distant
points than local field theory would allow.  Obviously, if our model is
to be a success we must resolve this point.

Before concluding we want to summarize the main points of our argument for
those readers who may have gotten lost in the details.  We have constructed
a theory of quantum cosmology based on the holographic principle and the UV/IR
correspondence of M-theory.   The very initial {\it Planck era} of 
the universe,
when the particle horizon size is of order the Planck scale does not 
have any sort
of geometrical description.  This merges into a phase dominated by a 
dense black hole
gas whose equation of state is $p=\rho$.  The geometrical
description of this phase by an observer in a single horizon volume is a flat
FRW universe filled with a black hole that saturates the FSB entropy bound.
Because the homogeneous $p=\rho$ fluid saturates the bound, we have 
solved the horizon/homogeneity problem.
There are no inhomogeneous perturbations of
these initial conditions.  Picturesquely we can say that this is a 
consequence of the fact that a black hole
represents an enormous number of states that
all have the same geometry.  Arguments based on the picture of a 
dense black hole gas (more properly,
just on the properties of holographic entropy formulae)
also explain why the FRW geometry must be locally flat.  Negatively curved
regions support much less entropy and positively curved regions
recollapse on the Hubble time scale and become small black holes embedded in
a flat background.  They are absorbed into the dense black hole gas.

The $p=\rho$ fluid is stable to small fluctuations, but in a
statistical ensemble of states compatible with the holographic
principle there is some small probability that there will be large
enough fluctuations to break away from the $p=\rho$ fluid.  We
argue that these form little bubbles of gas whose constituents are
small black holes, which quickly decay into radiation. The
probability $P_0$ for these fluctuations to occur, which
eventually determines the reheat temperature of the universe,
depends on physics during the Planck era, though we are able to
argue that it is small.   The same is true of the size of the
inhomogeneous fluctuations in the density of these dilute bubbles.
However, we are able to calculate the form of the spectrum of
these fluctuations in a certain range of wavelengths. It is a
Gaussian Harrison-Zeldovich spectrum.  The origin of its scale
invariance is the special scaling law for graviton correlation
functions in the $p=\rho$ fluid.  We have emphasized that this
result depends primarily on a conformal isometry of the $p=\rho$
FRW geometry, and may be more general than its field theoretic
derivation. This fluctuation formula is the most important result
of this paper. Unfortunately, in its current state the model also
seems to predict this scaling formula only for wavelengths much
smaller than the current Hubble scale. We hope that this
discrepancy is due to our invalid used of field theory to describe
fluctuations in the $p=\rho$ gas.

We then considered the monopole problem.  In M-theory, monopoles
are objects that appear only after we dimensionally reduce.  Some
of the details of monopole physics therefore depend on the
compactification to four dimensions. This is summarized in a
single new parameter $M_0$.  At early times we have an 11 (or
perhaps 10) dimensional fluid of black holes.  When the horizon
gets as large as the KK scale there is a Gregory-LaFlamme
transformation to a gas of wrapped black branes, which behave
like four dimensional black holes.  $M_0$ is the mass of these
black holes.   It is greater than $10^3$ in four dimensional
Planck units, perhaps significantly greater if the
compactification is highly anisotropic, like that of Horava and
Witten.  A small fraction of these black holes are magnetically
(and electrically) charged but as the black holes merge during
the $p=\rho$ phase the charge increases.  Finally, the volume of
$p=0$ regions becomes much larger than that of $p=\rho$ regions,
and the proper picture of the universe consists of huge
magnetically charged black holes which form a nonrelativistic
fluid. If the QED $\theta$ angle is nonzero, these also have a
fractional electric charge and all of their interactions,
including Hawking decay, should be CP violating.
  This picture persists until
Hawking radiation turns it into a radiation dominated universe
with a gas of highly charged extremal magnetic black holes
embedded in it. The requirement that monopole black holes not
dominate the energy density of the universe determines one
combination of the parameters $P_0$ and $M_0$. Consistency of
nucleosynthesis puts another constraint on these parameters. It
is nontrivial that both constraints can be satisfied.  This
result depends on the value of $M_0$, on the entropy formula for
charged black holes, and (very sensitively) on the value of the
electromagnetic coupling at the GUT scale. The reheat temperature
cannot be much higher than nucleosynthesis scale.

We also briefly discussed baryogenesis, gravitinos, and dark matter.
It appears that all of these can be safely accomodated in holographic cosmology
though there may be some constraints on the mechanism of SUSY breaking.
A particularly attractive and economical scenario has low energy SUSY breaking
which implies stable relativistic gravitinos, and uses a DFSK axion 
with GUT scale
decay constant as the dark matter.  Baryogenesis can be achieved with 
the AD mechanism.
We also suggested that the intrinsic CP and baryon number violation 
of charged black
hole decay might by itself explain the baryon asymmetry.

The relic black monopoles are the most unusual feature of our scenario.
Their number density is safely below existing limits, but might be substantial.
We suspect that in matter rich regions of the universe like our galaxy, they
may all have fallen long ago into the centers of gravitationally bound systems.
The discovery of even one such object would have profound implications for the
study of physics at extremely high energies.  Being pessimists by nature, we
fear that more refined versions of our calculations will show that 
the relic density
is too small to permit us to hope that they will be found in the near term.

It is clear that we have made only the roughest sketch of holographic 
cosmology in this paper.
Filling in factors of order one that may be raised to high powers is 
clearly in order.
One of the most interesting calculations suggested by our study is 
the generation of a particle
antiparticle asymmetry by dyonic
black holes in the presence of a $\theta$ angle.   The symmetries 
(rather the lack of them)
suggest that this is possible, but we do not understand the
precise mechanism by which it occurs, nor the order of magnitude of 
the asymmetry that is generated.
Other directions for study are the dependence
of the monopole physics on the compactification, the question of whether the
higher dimensional physics affects our picture in more substantial 
ways ({\it e.g}
altering the fluctuation spectrum at shorter but still accessible 
wavelengths) and
the interactions of axions and axion strings with the black monopoles
(we suspect that initially there may be axion strings passing through 
each monopole).
Finally, the importance and apparent simplicity of the $p=\rho$ fluid 
in our picture motivates
us to attempt to construct it in an intrinsically quantum manner 
using the formalism presented in our previous paper.
Perhaps this is the simple system that will enable us to figure out
some of the rules of the quantum theory of gravity.

\section{Acknowledgments}

We would like to thank Raphael Bousso, Michael Dine, Michael Douglas, 
Nemanja Kaloper, Bob Mcnees and Lenny Susskind for very useful
discussions.  W.F. would like to thank the members of the NHETC at
Rutgers University, for their hospitality while this paper was completed.

We would also like to thank the organizers of the Benasque Science Center
Conference, July 15-27, 2001, and the proprietors of cafes too 
numerous to mention
in the town of Benasque for their contributions to this paper.
The research of T.B was supported in part by DOE grant number 
DE-FG03-92ER40689, the
research of W.F. was supported in part by NSF grant- 0071512.

%


\end{document}